\documentclass[1p]{elsarticle}
\usepackage{graphicx} 
\usepackage{dcolumn}
\usepackage{bm}
\usepackage{amssymb}
\usepackage{amsmath,amsthm}
\usepackage{bm}
\usepackage{mathrsfs}

\usepackage{slashed}
\begin{document} 
\begin{frontmatter}
\title{$\mathcal{O}(\alpha^2)$ ISR effects with a full electroweak one-loop correction for a top pair-production at the ILC}
\author{Junpei Fujimoto$^{a}$, Yoshimasa Kurihara$^{a}$, Nhi M. U. Quach$^{a,b}$}
\address{$^b$ The Graduate University for Advanced Studies (SOKENDAI), Hayama, Kanagawa 240-0193, Japan. }
\address{$^a$ High Energy Accelerator Research Organization (KEK), Tsukuba, Ibaraki 305-0801, Japan. }
\begin{abstract}
Precise predictions for an $e^+e^-\rightarrow t\bar{t}$ cross-section are presented in the energy region from 400 GeV to 800 GeV.
Cross-sections are estimated including the beam-polarization effects with full $\mathcal{O}(\alpha)$, and also with the effects of the initial-state photon emission.
A radiator technique is used for the initial-state photon emission up to two-loop orders.
In this investigation, a weak correction is defined as the full electroweak corrections without the initial-state photonic corrections.
As a result, it is determined that the total cross-section of a top quark pair-production receives the weak corrections of $+4\%$ over the trivial initial state corrections at a center of mass energy of 500 GeV. 
Among the initial state contributions, a contribution from two-loop diagrams gives less than $0.11\%$ correction over the one-loop ones at the center of mass energies of from $400$ GeV to $800$ GeV.
In addition, the effect of a running coupling constant is also discussed.
\end{abstract}
\end{frontmatter}
%
%
\section{Introduction}
The standard theory of particle physics has been finally established after the discovery of the Higgs boson\cite{Aad:2012tfa,Chatrchyan201230} in 2012.
The next major challenge in particle physics is the search for a more fundamental theory beyond the standard model (BSM).
In this regard, the role of the Higgs boson and the top quark is considered to be critical.
Since the top quark is the heaviest fermion with a mass in the electroweak symmetry-breaking regime, it is naturally expected to have a special role in the BSM\@. 

The international linear collider\cite{Behnke:2013xla} (ILC), which is an electron-positron colliding experiment with a center of mass (CM) energies above 250 GeV, is proposed and intensively discussed as a future project of high-energy physics. 
One of the main goals of the ILC experiments is the precise measurement of the properties of the top quark.
Detailed Monte Carlo studies have shown that the ILC can measure most of the standard model parameters to within sub-percent levels\cite{Baer:2013cma}.
Theoretical predictions are required with a new level of precision because of the improvement in the experimental accuracy of the ILC.
In particular, a radiative correction due to the electroweak interaction (including spin polarizations) is mandatory based on this requirement.
Before the discovery of the top quark, a full electroweak radiative correction was performed for an $e^-e^+\rightarrow t\bar{t}$ process at a lower energy\cite{Fujimoto:1987hu,BEENAKKER199124}, and was then independently obtained for higher energies\cite{Fleischer:2002rn,Fleischer2003}.
The same correction for a $e^-e^+\rightarrow t\bar{t}\gamma$ process has also been reported\cite{Khiem2013}.
NLO QCD corrections for on-shell $t\bar{t}$ and $t\bar{t}H$ including decays are calculated in Ref.\cite{Nejad:2016bci}.
A detailed study of the electroweak correction for top quark decay is also reported in Refs.\cite{Oliveira:2001vw,Oggero:2014yfa}.
A possible search of the minimum SUSY particles based on loop corrections of the top-quark pair-production at the ILC is reported in \cite{doi:10.1093/ptep/ptx048}.

Recently, electroweak radiative corrections for the process 
$e^-e^+\rightarrow t \bar{t}$ $\rightarrow b \bar{b} \mu^+\mu^-\nu_{\mu} \bar{\nu}_{\mu}$ including the spin-polarization effects have been reported\cite{Quach:2017ijt} by the authors of a present report.
Unfortunately, a complete electroweak correction for the six-body final-state is impossible based on the current limits of computing power because the number of Feynman diagrams concerned is too large.
In Ref.\cite{Quach:2017ijt}, authors used a simple narrow width approximation (NWA) for the top-quark production and decay. 
A more precise method to treat both production and decay at a one loop-order is the double-pole approximation method.
This method was initially developed for a $W$-boson pair production\cite{Denner:2000bj, Accomando:2004de}, and was subsequently applied to top quark production\cite{Denner:2016jyo}.

Among several sources of radiative correction, it is known that the initial state photonic correction (ISR) accounts for the largest contribution in general, and thus, it is very important for the precise estimation of production cross-sections.
Top-quark decay is not treated in this study, because the ISR effect on the total cross-section does not strongly depend on its decay process.
In this report, the precise estimation of the effect due to the initial-state photon-radiation is discussed in detail.

%
%
\section{Calculation Method}\label{CM}
\subsection{GRACE system}
For precise cross-section calculations of the target process, a GRACE-Loop system is used in this study.
The GRACE system is an automatic system for calculating the cross-sections of scattering processes at a one-loop level for the standard theory\cite{Belanger2006117} and the minimum SUSY model\cite{PhysRevD.75.113002}. 
The GRACE system has been used to treat electroweak processes consisting of two, three or four particles in the final state \cite{Belanger2003152,Belanger2003163, Belanger2003353,Kato:2005iw} at the one-loop order.
The renormalization of the electroweak interaction is performed using on-shell scheme\cite{doi:10.1143/PTPS.73.1,doi:10.1143/PTPS.100.1}. 
Divergences in the infrared limit are regulated using fictitious photon-mass\cite{doi:10.1143/PTPS.100.1}.
The symbolic manipulation package FORM\cite{Vermaseren:2000nd} is used to deal with all Dirac and tensor algebra in $n$-dimensions. 
For loop integrations, all tensor one-loop integrals are reduced to scalar integrals using our own formalism\cite{Belanger2006117}, then performed integrations using packages FF\cite{vanOldenborgh:1990yc} or LoopTools\cite{Hahn:1998yk}. 
Phase-space integrations are performed using an adaptive Monte Carlo integration package BASES\cite{Kawabata1986127,KAWABATA1995309}.
For numerical calculations, we used a quartic precision for floating variables. 

While using $R_\xi$-gauge for the linear gauge-fixing terms in the GRACE system, the non-linear gauge fixing Lagrangian\cite{Boudjema:1995cb,Belanger2006117} is also employed to check the system.
Numerical tests were performed to confirm that the amplitudes are independent of all redundant parameters at approximately $20$ digits at several randomly chosen phase points before calculating the cross-sections.
In addition to the aforementioned checks, the soft-photon cut-off independence was examined:
cross-sections at the one-loop level, results must be independent of the head-photon cut-off parameter $k_c$. 
We confirmed that, while varying a parameter $k_c$ from $10^{-4}$ GeV to $10^{-1}$ GeV,  the results of numerical phase-space integration are consistent with each other within the statistical errors of numerical integrations, which is typically on the order of $0.1\%$.

\subsection{Radiator method}
The effect of the initial photon emission can be factorized when the total energy of the emitted photons is sufficiently small compared to the beam energy or the small angle (co-linear) emission.
The calculations under such an approximation are referred to as the ``soft-colinear photon approximation(SPA)''.
Using the SPA, the corrected cross-sections with the initial state photon radiation(ISR), $\sigma_{ISR}$, can be obtained from the tree cross-sections $\sigma_{Tree}$ using a structure function $H(x,s)$ as follows:
\begin{eqnarray} 
\sigma_{ISR}&=&\int^1_0 dx~H(x,s)\sigma_{Tree}\left(s(1-x)\right),\label{ISRTree}
\end{eqnarray}
where $s$ is the CM energy squared and $x$ is the energy fraction of an emitted photon.
The structure function can be calculated using the perturbative method with the SPA.
Concrete formulae of the structure function are calculated up to two loop orders\cite{doi:10.1143/PTPS.100.1}. 
A further improvement of the cross-section estimation is possible using the ``exponentiation method''.
For initial state photon emissions under the SPA, the probability of emitting each photon should be independent of each other.
Thus, the probability of emitting any number of photons can be calculated as;
\begin{eqnarray} 
p&=&\sum_{k=0}^{\infty}p_k=\sum_{k=0}^{\infty}\frac{1}{k!}(p_1)^k,
\end{eqnarray}
where $p_k$ is the probability of emitting $k$ photons. 
A factor $1/k!$ is necessary due to the appearance of $k$ identical particles (photons)in the final state.
This is essentially a Taylor expansion of the exponential function.
Therefore, the effect of multiple photon emissions can be estimated by making the one-photon emission probability the argument of the exponential function.
This technique is referred to as to the exponentiation method.
 When the exponentiation method is applied to the cross-section calculations at loop level, the corrected cross-sections cannot simply be expressed as the formula (\ref{ISRTree}), because the same loop corrections are included in both the structure function and loop amplitudes. 
To avoid a double counting of the same corrections in the structure function and loop amplitudes, the terms of corrections have to be rearranged.

The total cross-section at one-loop (fixed) order without the exponentiation, which is denoted as $\sigma_{NLO;fixed}$, can be expressed as;
\begin{eqnarray} 
\sigma_{NLO;fixed}&=&\sigma_{Loop}+\sigma_{Soft}+\sigma_{Hard}+\sigma_{Tree},\label{sNLOfinx}
\end{eqnarray}
where $\sigma_{Loop}$, $\sigma_{Soft}$ and $\sigma_{Hard}$ are the cross-sections from the loop diagram, and the soft and hard real-emission corrections, respectively.
A photon with energy that is greater (less) than the threshold energy $k_c$ is defined as a hard (soft) photon, respectively.
The soft-photon cross-section can be expressed as $\sigma_{Soft}=\sigma_{Tree}\delta_{SPA}$, where a factorized function $\delta_{SPA}$ is obtained from the real-radiation diagrams using the SPA.
The SPA consists of three parts, the initial-state, final-state radiations, and their interference terms. 
For the initial state radiation, it can be written as;
\begin{eqnarray}
\delta_{SPA}&=&\frac{\alpha}{\pi}\left(
2\left(L-1\right)\log{\frac{m_e}{\lambda}}+\frac{1}{2}L^2-2l\left(L-1\right)
\right),
\end{eqnarray}
where $\alpha$, $m_e$ and $\lambda$ are the fine structure constant, electron mass and fictitious photon mass, respectively.
Here two large log-factors appear as $L=\log{(s/m_e^2)}$ and $l=\log{(E/k_c)}$, where $E$ is a beam energy and $s=4E^2$.
The threshold energy $k_c$ is included in both $\sigma_{Hard}$ and $\delta_{SPA}$, and the total cross-section must be independent of the value of $k_c$ after summing up all contributions.
At the same time, final results are also independent of the photon mass $\lambda$ due to the cancellation among $\lambda$ contributions in $\sigma_{Loop}$ and $\delta_{SPA}$. 

The cross-section with a fixed order correction  (\ref{sNLOfinx}) can be improved using the exponentiation method.
To avoid double counting of the terms that appear in both the loop corrections and the structure function, the terms must be re-arranged as follows:
\begin{eqnarray} 
\sigma_{NLO;ISR}&=&
\left(\sigma_{Loop}-\sigma_{Tree}\delta_{ISL}\right)+\widetilde{\sigma}_{Soft}
+\sigma_{Hard}+\widetilde{\sigma}_{ISR},\label{sNLOISR}
\end{eqnarray}
where $\delta_{ISL}$ is a correction factor from the initial state photon-loop diagrams, that can be given as;
\begin{eqnarray} 
\delta_{ISL}&=&\frac{2\alpha}{\pi}\left(
-\left(L-1\right)\log{\frac{m_e}{\lambda}}-\frac{1}{4}L^2+\frac{3}{4}L+\frac{\pi^2}{3}-1
\right).
\end{eqnarray}
This term must be subtracted from $\sigma_{Loop}$ because the same contribution is also included in the structure function $H(s,x)$.
A term $\widetilde{\sigma}_{Soft}$ includes only the final-state radiation and interference terms between the initial and final state radiations.
Instead, $\widetilde{\sigma}_{ISR}$ gives the improved cross-section including the initial-state radiation using the radiator method\cite{doi:10.1143/PTPS.100.1}.
The total cross-section can be calculated using the radiator function as;
\begin{eqnarray} 
\sigma_{ISR}&=&\int^{k^2_c/s}_0 dx_1 \int^{1-x1}_0dx_2~D(x_1,s)D(x_2,s)\sigma_{Tree}\left(sx_1x_2)\right).
\end{eqnarray}
The radiator function $D(x,s)$, which corresponds to the square root of the structure function,  gives the probability of emission of a photon with an energy fraction of $x$ at the CM energy square $s$.
In this method, electrons and positrons can emit different energies, and thus a finite boost of the CM system can be treated.
The radiator function can be obtained as\cite{doi:10.1143/PTPS.100.1}.
\begin{eqnarray} 
D(1-x,s)^2&=&H(x,s)=\Delta_2\beta x^{\beta-1}
-\Delta_1\beta\left(1-\frac{x}{2}\right)\nonumber\\
&~&+\frac{\beta^2}{8}\left[
-4(2-x)\log{x}-\frac{1+3(1-x)^2}{x}\log{(1-x)}-2x
\right],\label{ISRLoop}
\end{eqnarray}
where
\begin{eqnarray*}
\beta&=&\frac{2\alpha}{\pi}\left(\log{\frac{s}{m_e^2}}-1\right),\\
\Delta_2=1+\delta_1+\delta_2,&~&\Delta_1=1+\delta_1\\
\delta_1=\frac{\alpha}{\pi}\left(\frac{3}{2}L+\frac{\pi^2}{3}-2\right),&~&
\delta_2=\left(\frac{\alpha L}{\pi}\right)^2
\left(
-\frac{1}{18}L+\frac{119}{72}-\frac{\pi^2}{3}
\right).
\end{eqnarray*}
In this case, the photon mass $\lambda$ is cancelled between $\sigma_{Loop}$ and $\sigma_{Tree}\delta_{ISL}$, and the threshold energy $k_c$ is cancelled between $\sigma_{Hard}$ and $\widetilde{\sigma}_{ISR}$.
This result is obtained based on perturbative calculations for initial-state photon emission diagrams up to two-loop orders \cite{doi:10.1143/PTPS.100.1}.
Terms with $\alpha^2$ in (\ref{ISRLoop}) are obtained from the two-loop diagrams.

%
%
\section{Results and discussions}\label{R&D}
\subsection{Input parameters}
The input parameters used in this report are listed in Table \ref{parameters}. 
The mass of the light quarks (i.e., other than the top quark) and $W$ boson are chosen to be consistent with low-energy experiments\cite{Khiem2015192}.
Other particle masses are taken from recent measurements\cite{Olive:2016xmw}.
The weak mixing-angle is given using the on-shell condition $\sin^2{\theta_W} = 1- m_W^2/m_Z^2$ because of our renormalization scheme.
The fine-structure constant $\alpha=1/137.0359859$ is obtained from the low-energy limit of Thomson scattering, because of the renormalization scheme used.

All cross-section calculations in this report are performed with $100\%$ left (right) polarization of an electron (positron), respectively, because the initial state of the photonic corrections are independent of the beam polarization
	\begin{table}[b]
		\begin{center}
			\begin{tabular}{|c|c||c|c|}
\hline
$u$-quark mass & $58.0\times10^{-3}$ GeV & $d$-quark mass & $58.0\times10^{-3}$ GeV \\
$c$-quark mass & $1.5$ GeV & $s$-quark mass & $92.0\times10^{-3}$ GeV \\
$t$-quark mass & $173.5$ GeV & $b$-quark mass & $4.7$ GeV \\
$Z$-boson mass & $91.187$ GeV & $W$-boson mass & $80.370$ GeV \\
Higgs mass & 126 GeV & ~ & ~\\
\hline
			\end{tabular}
			\caption{Particle masses used in analysis.}
\label{parameters}
		\end{center}
	\end{table}

\subsection{Electroweak radiative corrections}

\subsubsection{total cross-sections}
\begin{figure}[t]
  	\begin{center}
  		\includegraphics[width={8cm}]{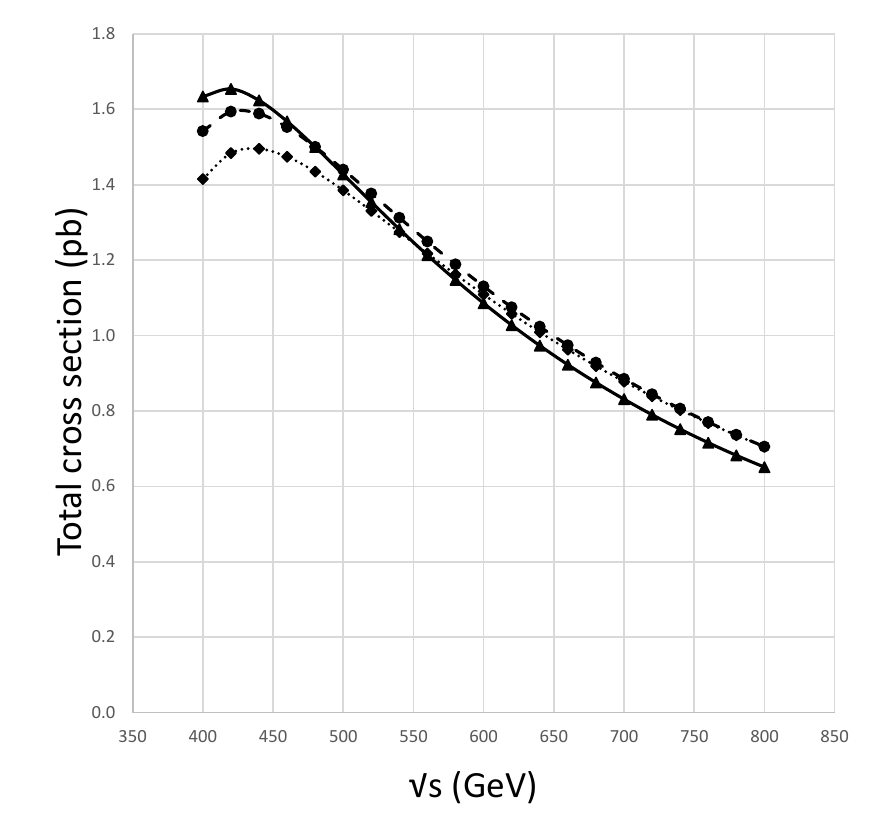}
  	\end{center}
  	\caption{
  	 Total cross-sections of top quark pair production at the tree level, tree with the ISR and one-loop with the ISR corrections are shown as a function of the CM energy.
The solid line (with triangle points) and dashed line (with circle points) show the tree and NLO cross-sections with the ISR.
The dotted line (with rectangle points) shows the cross-sections with only the ISR correction on the tree cross-sections.
}
  	\label{fig1}
  \end{figure}

At first, the fixed order correction without using an exponentiation method is investigated. 
The total cross-sections obtained at leading (tree) and next-to-leading order (NLO) calculations are shown in Figure~\ref{fig1}.
These results are the same as those included in our previous report\cite{Quach:2017ijt} for a simple NLO correction.
The consistency between our current and previous results\cite{Fujimoto:1987hu,Fleischer:2002rn,Fleischer2003}was numerically confirmed after adjustment of the input parameters.
The NLO calculations near the top-quark production threshold (at an approximate CM energy of $400$ GeV) reveals negative corrections of approximately $7\%$.
The corrections become very small in the vicinity of the CM energy of $500$ GeV and increase in the high-energy region to $2.8\%$ at the CM energy of $800$ GeV.
Considering several types of radiative corrections, e.g., the initial and final state photon radiation, the vertex and box correction, etc., the initial-state photonic correction gives the largest contribution at the high energy region.
As shown in Figure~\ref{fig1}, the cross-sections at the tree level including the ISR correction (a dotted line in the figure) are almost the same as the full order $\mathcal{O}(\alpha)$ electroweak correction (a dashed line in the figure) at CM energies above $700$ GeV.
This implies that the main contribution of the higher order corrections is caused by the initial state photonic corrections.
On the other hand, other corrections from the loop diagrams also give a large correction near the threshold region.
At a CM energy of approximately $500$ GeV, these effects are accidentally canceled and result in a small correction of the total cross-section.

\begin{figure}[t]
  	\begin{center}
  		\includegraphics[width={8cm}]{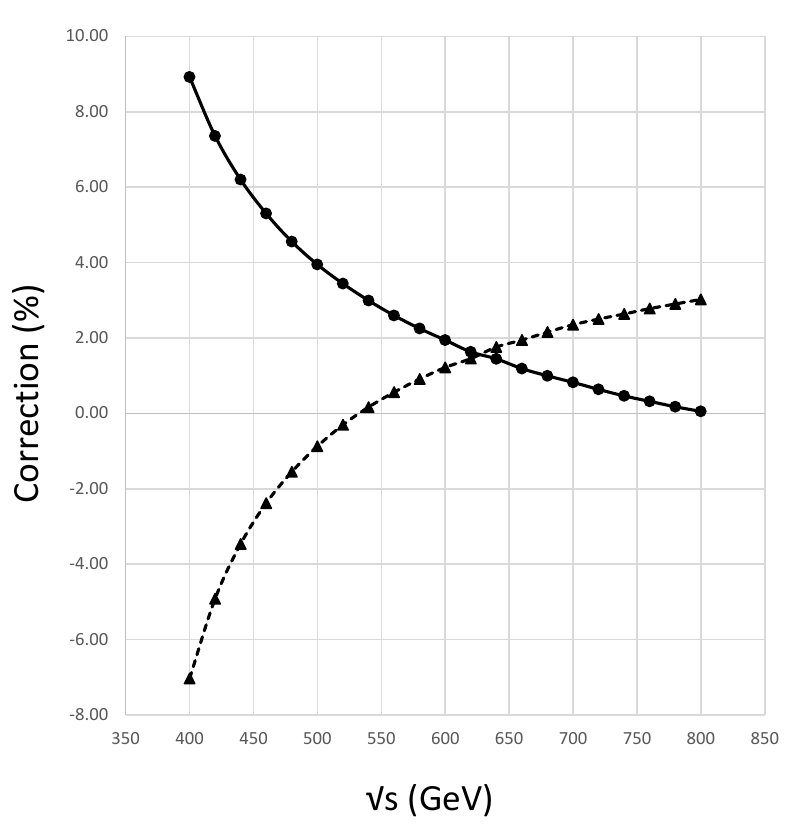}
  	\end{center}
  	\caption{
  	The degree of correction for the full $\mathcal{O}(\alpha)$ electroweak radiative corrections (dashed line with triangle points) and the weak corrections (solid line with circle points).
  	}
  	\label{fig2}
  \end{figure}

After subtracting a trivial ISR correction from the total corrections, the``pure'' weak correction in the full $\mathcal{O}(\alpha)$ electroweak radiative corrections can be examined.
The NLO correction degree is defined as:
\begin{eqnarray}
\delta_{NLO}&=&\frac{\sigma_{NLO;fixed}-\sigma_{Tree}}{\sigma_{Tree}},
\end{eqnarray}
the weak correction degree is defined as:
\begin{eqnarray}
\delta_{weak}&=&\frac{\sigma_{NLO;ISR}-\sigma_{ISR}}{\sigma_{ISR}}.
\end{eqnarray}
In the definition of $\delta_{weak}$, the trivial initial-state photonic corrections are subtracted from the full $\mathcal{O}(\alpha)$ electroweak radiative corrections, and thus, $\delta_{weak}$ shows a fraction of the mainly weak-correction in the full $\mathcal{O}(\alpha)$ electroweak radiative corrections.
The behaviors of $\delta_{weak}$ and $\delta_{NLO}$ are shown in Figure~\ref{fig2} with respect to the CM energies.
It can be seen that the weak correction becomes progressively smaller in a high energy region, and has a value of almost zero at the CM energy of $800$ GeV.
However, at the CM energy of $500$ GeV, the pure-weak corrections gives $+4\%$ correction over the trivial ISR corrections.

\subsubsection{angular distribution}
  \begin{figure}[t]
  	\begin{center}
  		\includegraphics[width={8cm}]{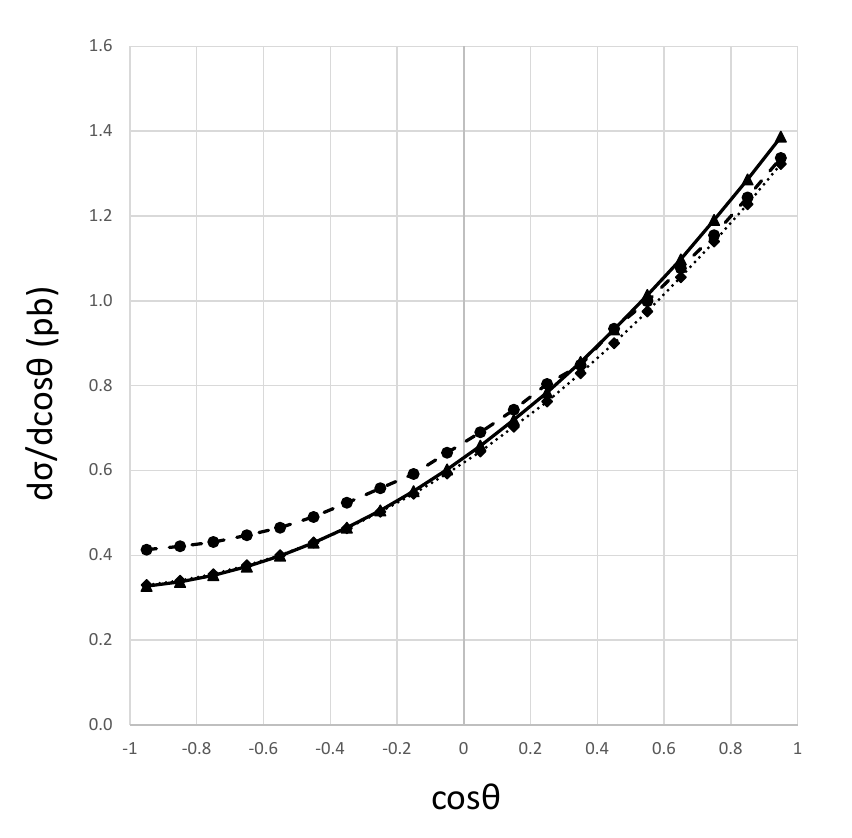}
  	\end{center}
  	\caption{
Angular distributions of a top quark.
A solid line (with triangle points), dotted line (with rectangle points) and dashed line (with circle points) show the tree, tree with ISR and NLO with ISR cross-sections.
  	}
  	\label{fig3}
  \end{figure}
As previously indicated, the radiative correction does not accidentally change the total cross-section at the CM energy $500$ GeV.
Although the total cross-section remains the same after the radiative correction, this is not the case for the angular distribution.
In general, the real-photon emission affects the angular distribution such that a steep peak is less pronounced. 
In reality, the ISR correction (a dotted line with rectangle points) reduces the forward peak of a top quark production at a tree level (a solid line with triangle points) as shown in Figure \ref{fig3}.
In addition, the weak correction increases the backward scattering as shown by a dashed line with circle points in Figure \ref{fig3}.
As a result, the total cross-section does not change significantly.

\subsection{Photonic correction at two-loop order}
\begin{figure}[t]
  	\begin{center}
  		\includegraphics[width={8cm}]{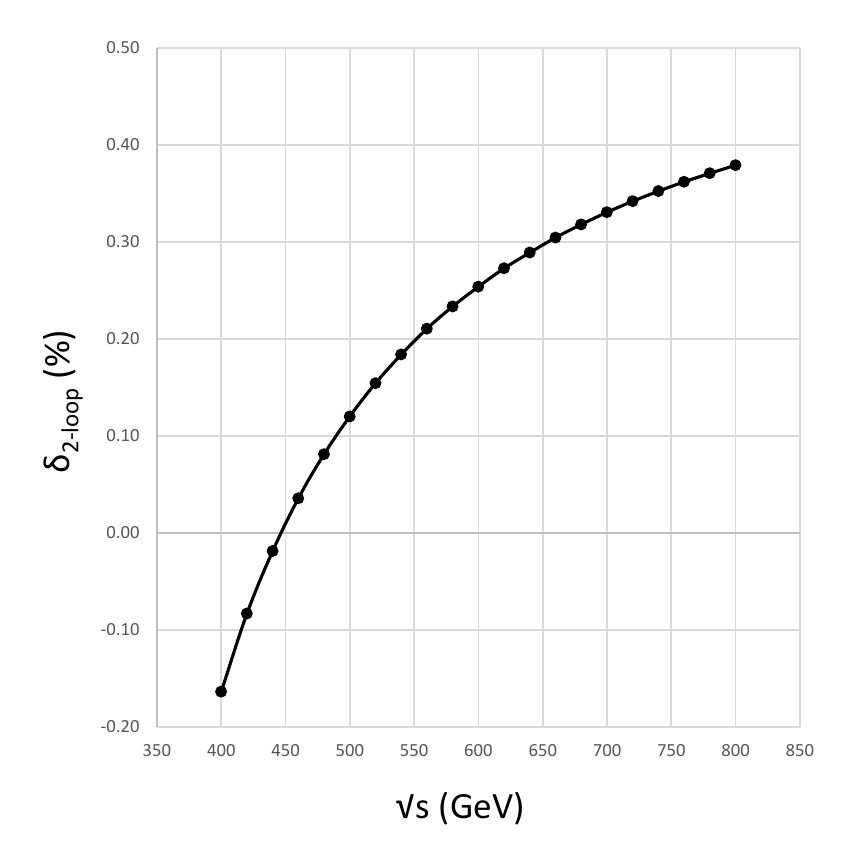}
  	\end{center}
  	\caption{
  	The contribution of the two-loop diagrams to the ISR corrections are show as a function of the CM energies.
  	The definition of  $\delta_{2\rm{-} loop}$ is given in (\ref{d2l}).
  	}
  	\label{fig4}
  \end{figure}

The structure function $H(s,x)$ given in (\ref{ISRLoop}) include two-loop effects.
All of the aforementioned results were obtained using a full formula (\ref{ISRLoop}).
As mentioned in the previous subsection,  a main contribution of the radiative corrections comes from the initial state photonic-correction in the high energy region.
Therefore, the ISR correction is one of many important terms of the radiative corrections associated with these energies.
If the two-loop contribution has a significant fraction in the full correction, even higher-loop corrections must be considered in future experiments.
The fraction of two-loop contribution over the one-loop one is defined as
\begin{eqnarray}
\delta_{2-loop}&=&\frac{\sigma_{ISR}-\sigma^{(1)}_{ISR}}{\sigma^{(1)}_{ISR}},\label{d2l}
\end{eqnarray}
where $\sigma^{(1)}_{ISR}$ shows the ISR corrected cross-sections using the structure function (\ref{ISRLoop}) and omitting $\delta_2$ and $\beta^2$ terms.
Numerical results are shown in Figure~\ref{fig4}.
The two-loop contribution is smaller than $0.4\%$ in the energy region between $400$ GeV to $800$ GeV as shown in Figure~\ref{fig4}.

\subsection{Running coupling}
The running coupling method is another improvement of the cross-section estimation.
This method also takes higher order effects of vacuum polarization diagrams into account as an effective coupling constant.
After summing up the contributions from fermion-loops on gauge-boson propagators, they solely form a gauge-invariant subset, an electro-weak coupling effectively varies according to the four-momentum square of the propagators.
The effective coupling $\alpha(|q^2|)$ at the energy scale $|q^2|$ due to a fermion with unit charge can be written as;
\begin{eqnarray}
\alpha\left(|q^2|\right)&=&\frac{\alpha\left(\mu^2\right)}
{1-\frac{\alpha\left(\mu^2\right)}{3\pi}\log{\left(\frac{|q^2|}{\mu^2}\right)}},\label{12}
\end{eqnarray}
where $q^2$ is the four-momentum square of a propagator of a target process.
Contributions from all leptons and quarks except a top quark are summed in calculations. 
At the same time, the weak mixing angle $\theta_W$ is obtained from the Fermi weak-coupling constant $G_F$ as\cite{Altarelli:473529};
\begin{eqnarray}
\sin^2{\theta_W}&=&
\frac{\pi\alpha\left(|q^2|\right)}{\sqrt{2}G_Fm_W^2}.\label{13}
\end{eqnarray}
In this study, $q^2=s'$ is used for $tt\gamma$ and $ttZ$ vertices, where $s'$ is an effective CM energy-square after ISR photon emission.
For calculations including top decays, a coupling constant for $qqW$ vertices is provided using eqs. (\ref{12}) and (\ref{13}) with $q^2=m_W^2$.   

\begin{figure}[t]
  	\begin{center}
  		\includegraphics[width={8cm}]{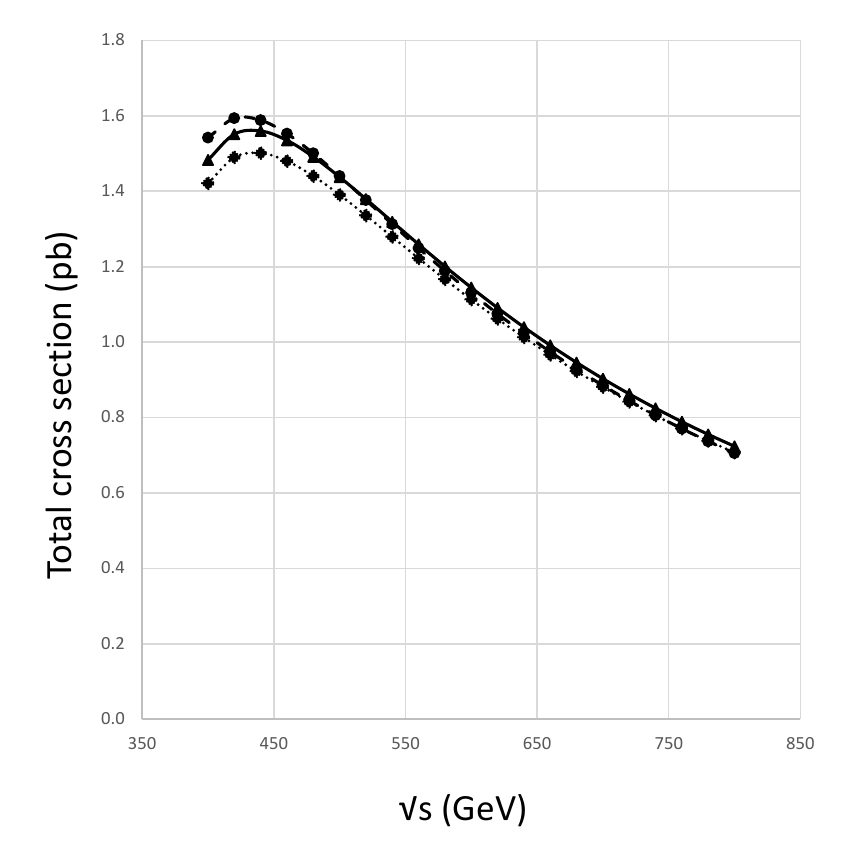}
  	\end{center}
  	\caption{
Total cross-sections of top quark pair production of the tree and NLO cross-sections with the ISR correction, and the improved Born cross-sections are shown as a function of the CM energy.
The dotted line (with rectangle points) and dashed line (with circle points) show the tree and NLO cross-sections with the ISR.
The solid line (with triangle points) shows the cross-sections of the improved Born approximation.
  	}
  	\label{fig5}
  \end{figure}

The improved cross-section $\sigma_{imp}$ can be written as;
\begin{eqnarray}
\sigma_{imp}&=&\int^1_0 dx_1\int^{1-x_1}_0 dx_2~D(x_1,s)D(x_2,s)\sigma_{Tree}\left(\alpha\left(sx_1x_2\right);sx_1x_2\right),
\end{eqnarray}
where $\sigma_{Tree}\left(\alpha(q^2);q^2\right)$ is the tree cross-section at the CM energy-square $q^2$ with the coupling constant $\alpha(q^2)$.
This method is referred to as the improved Born approximation (IBA).
Numerical results based on the IBA are summarized and compared with the proposed method in Figure \ref{fig5}.
Here, the measured value of $\alpha(\mu^2=(2m_W)^2)=128.07$\cite{Altarelli:473529} is used. 
Above the CM energy of $500$ GeV, the IBA yields approximated values that are better than $2\%$ with respect to the NLO cross-sections with the ISR.
The IBA is including a running effect of a coupling constant due to a contribution from vacuum polarization diagrams of a gauge boson exchange.
The difference between the IBA and the proposed method is induced by box-diagrams including W-bosons, which are not included in the IBA method.
For future precision measurements at the ILC, a precision at the sub-percent level can be expected, and thus, the simple IBA is not sufficiently precise for measurement acquisition at the ILC.
%
%
\section{Summary}\label{Summary}
We calculated the precise cross-sections of an $e^+e^-\rightarrow t\bar{t}$ process in the energy region from 400 GeV to 800 GeV.
In particular, the initial-state photon emissions were discussed in detail. 
An exponentiation technique was applied for the initial-state photon emissions up to two-loop orders.
It was determined that the total cross-section of a top quark pair-production at a center of mass energy of 500 GeV receives the weak corrections of $+4\%$ over the trivial ISR corrections.
Among the ISR contributions, two-loop diagrams resulted in less than $0.4\%$ correction with respect to the one-loop ones at the CM energies from $400$ GeV to $800$ GeV.

The improved Born approximation yielded cross-sections better than $2\%$ compared with the NLO cross-sections with ISR.

\section*{Acknowledgement}
I would like to thank prof.~T.~Kaneko and prof.~F.~Yuasa for their continuous encouragement and fruitful discussion.
We would like to thank Editage for English language editing.

%
\bibliography{tT_ISR_revised_R3}
\end{document}